\documentstyle[amssymb,aps,preprint]{revtex}

\tightenlines

\begin{document}
\title{Supercooled vortex liquid and quantitative theory of melting of the
flux line lattice in type II superconductors. }
\author{$^{1}$Dingping Li\thanks{%
e-mail: lidp@mono1.math.nctu.edu.tw} and $^{2}$Baruch Rosenstein\thanks{%
e-mail: baruch@vortex1.ep.nctu.edu.tw}}
\address{$^{1}${\it National Laboratory of Solid State Microstructures and
Department of Physics, Nanjing University, Nanjing 210093, China}\\
$^{2}${\it National Center for Theoretical Sciences and} \\
{\it Electrophysics Department, National Chiao Tung University,Hsinchu }%
30050, {\it Taiwan, R. O. C}.}
\date{\today }
\maketitle

\begin{abstract}
A metastable homogeneous state exists down to zero temperature in systems of
repelling objects. Zero ''fluctuation temperature''\ liquid state therefore
serves as a (pseudo) ''fixed point''\ controlling the properties of vortex
liquid below and even around melting point. There exists Madelung constant
for the liquid in the limit of zero temperature which is higher than that of
the solid by an amount approximately equal to the latent heat of melting.
This picture is supported by an exactly solvable large $N$ Ginzburg - Landau
model in magnetic field. Based on this understanding we apply Borel - Pade
resummation technique to develop a theory of the vortex liquid in type II
superconductors. Applicability of the effective lowest Landau level model is
discussed and corrections due to higher levels is calculated. Combined with
previous quantitative description of the vortex solid the melting line is
located. Magnetization, entropy and specific heat jumps along it are
calculated. The magnetization of liquid is larger than that of solid by $%
1.8\%$ irrespective of the melting temperature. We compare the result with
experiments on high $T_{c}$ cuprates $YBa_{2}Cu_{3}O_{7}$, $DyBCO$, low $%
T_{c}$ material $(K,Ba)BiO_{3}$ and with Monte Carlo simulations.
\end{abstract}

\vskip 0.5cm
\flushleft{PACS numbers: 74.60.-w, 74.40.+k,  74.25.Ha,
74.25.Dw}

\newpage

\section{Introduction and the main idea}

Abrikosov flux lines (vortices) created by magnetic field in type II
superconductors strongly interact with each other creating highly correlated
configurations like the vortex lattice. In high $T_{c}$ cuprates thermal
fluctuations at relatively large temperatures are strong enough to melt the
lattice. Several remarkable experiments demonstrated that the vortex lattice
melting in high $T_{c}$ superconductors is first order with magnetization
jumps \cite{Zeldov} and spikes in specific heat \cite{Schilling}.
Magnetization and entropy jumps were measured using local Hall probes \cite%
{Zeldov}, SQUID \cite{Pastoriza,Welp}, torque magnetometry \cite%
{Willemin,Nishizaki} and integrating the specific heat spike \cite%
{Schilling,Bouquet}. It was found that in addition to the spike there is
also a jump in specific heat in $YBCO$ which was measured as well \cite%
{Schilling,Bouquet,Roulin}. These precise measurements pose a question of
accurate quantitative theoretical description of thermal fluctuations in
vortex matter.

The melting line in high $T_{c}$ cuprates has been studied mainly not very
far from $T_{c}.$ In this part of the phase diagram the Ginzburg - Landau
(GL) approach is generally appropriate to describe thermal fluctuations near
$T_{c}$ \cite{Tinkham,Blatter}. The GL model is however highly nontrivial
even within the lowest Landau level (LLL) approximation valid at relatively
high fields. This simplified model has only one parameter: the dimensionless
LLL scaled temperature $a_{T}$ $\thicksim $ $(T-T_{mf}(H))/(TH)^{2/3}$
(defined precisely in eq.(\ref{athl}) below). Over\ the last twenty years
great variety of theoretical methods were applied to study this model.
Brezin, Nelson and Thiaville \cite{Brezin} applied the renormalization group
(RG) method on the one loop level description of the vortex liquid. No
nontrivial fixed points of the (functional) RG equations were found and they
concluded therefore that the transition from liquid to the solid is first
order \cite{MooreRG}. Another often used approach applicable also beyond the
range of validity of GL model is to use elasticity theory description of the
vortex lattice and Lindermann criterion to determine the location of melting
line \cite{Mikitik}. However all those approaches do not provide a
quantitative theory of melting since these are one phase theories and in
order, for example, to calculate discontinuities at first order transition
accurate description of both phases is necessary..

Two perturbative approaches were developed and greatly improved recently to
describe both the solid and the liquid phases in the LLL GL model. The
perturbative approach on the liquid side was pioneered long ago by Thouless
and Ruggeri \cite{Ruggeri}. They developed a perturbative expansion around a
homogeneous (liquid) state in which all the ''bubble''\ diagrams are
resumed. Unfortunately they found that the series are asymptotic and
although first few terms provide accurate results at very high temperatures,
the series become inapplicable for $a_{T}$ less than $-2$, which is quite
far above the melting line (believed to be located around $a_{T}\sim -10$).
We recently obtained the optimized Gaussian series \cite{LiPRL} which are
convergent rather than asymptotic with radius of convergence of $a_{T}=-5$
still unfortunately above the melting point.

On the solid side, long time ago Eilenberger and Maki and Takayama \cite%
{Eilenberger} calculated the fluctuations spectrum around Abrikosov's mean
field solution. They noticed that the vortex lattice phonon modes are softer
than that of the acoustic phonons in atomic crystals and this leads to
infrared (IR) divergences in certain quantities. This was initially
interpreted as ''destruction of the vortex solid by thermal fluctuations''\
and the perturbation theory was abandoned. However the divergences look
suspiciously similar to ''spurious''\ IR divergences in the critical
phenomena theory and recently it was shown that all these IR divergences
cancel in physical quantities \cite{Rosenstein}. The series therefore are
reliable, and were extended to two loops, so that the LLL GL theory on the
solid side is now precise enough even around melting point.

Therefore the missing part is a theory in the region $-10<a_{T}<-5$ on the
liquid side. Moreover this theory should be very precise since free energies
of solid and liquid happen to differ only by few percents around melting.
Closely related to melting is the problem of nature of the metastable phases
of the theory. While it is clear that the overheated solid becomes unstable
at some finite temperature, it not generally clear whether the overcooled
liquid becomes unstable at some finite temperature (like water and other
molecular liquids, which however has a crucial attractive component of the
intermolecular force) or exists all the way down to $T=0$ as a metastable
state. The Gaussian (Hartree - Fock) variational calculation, although
perhaps of a limited precision, is usually a very good guide as far as
qualitative features of the phase diagram are concerned. Such a calculation
in the liquid was performed long ago \cite{Ruggeri}, while a significantly
more complicated one sampling also inhomogeneous states (vortex lattice) was
obtained recently \cite{Lirapid,LiSolid}. The gaussian results are as
follows. The free energy of the solid state is lower than that of the liquid
for all temperatures lower than melting temperature $a_{T}^{m}$. The solid
state is therefore the stable one below $a_{T}^{m}$, becomes metastable at
somewhat higher temperatures and is destabilized at $a_{T}=-5$. The liquid
state becomes metastable below the melting temperature, but unlike the
solid, does not loose metastability all the way down to $a_{T}^{m}=-\infty $
($T=0$). The excitation energy of the supercooled liquid approaches zero as
a power $\varepsilon \sim 1/a_{T}^{2}$. This general picture is supported in
section III by an exactly solvable large $N$ Ginzburg - Landau model of
vortex matter in type II superconductors.

Meantime in different area of physics similar qualitative results were
obtained. It was shown by variety of analytical and numerical methods that
liquid (gas) phase of the classical one component Coulomb plasma exists as a
metastable state down to very low temperature possibly $T=0$ \cite{Carvalho}%
. The quantum one component plasma - electron gas also shows similar feature %
\cite{Tanatar}. One considers quantum fluctuations instead of the thermal
ones (an analog of inverse temperature is coupling $r_{s}$, see section
III). It seems plausible to speculate that the same phenomenon would happen
in any system of point like or line like objects interacting via purely
repulsive forces. In fact the vortices in the London approximation are a
sort of repelling lines with the force even more long range than Coulombic.
This was an additional strong motivation to consider the above scenario in
vortex matter. In section III we provide both theoretical and
phenomenological evidence that the above scenario is the correct one.

Assuming absence of singularities on the liquid branch allows to develop an
essentially precise theory of the LLL GL model in vortex liquid (even
including supercooled liquid) using methods of theory of critical phenomena%
\cite{Amit,Baker}. The generally effective mathematical tool to approach a
nontrivial fixed point (in our case at zero temperature) is the Borel - Pade
(BP) transformation \cite{Baker}. Before embarking on this program in
following sections, we address several subtleties which prevented the use
and acceptance of the BP method in the past and then combine it with the
recently developed LLL theory of solids to calculate the melting line and
the magnetization and the specific heat jumps across the line. Very early on
Ruggeri and Thouless \cite{Ruggeri} tried to use BP (unfortunately a
"constrained" one, so that it interpolates smoothly with the solid,
assumption we believe is incorrect) to calculate the specific heat without
much success. It was shown by Wilkin and Moore \cite{Wilkin} that the
constrained BP does not converge, while the results for unconstrained BP
were inconclusive. They attributed this to the limited order of expansion
known at that time. Subsequent attempts to use BP for the calculation of the
melting line using longer series also ran into problems. Hikami, Fujita and
Larkin \cite{Hikami} tried to find the melting point by comparing the BP
energy with the one loop solid energy and obtained $a_{T}=-7$. However their
one loop solid energy was incorrect and, in any case, it was not precise
enough (as will become clear below the two loop contribution cannot be
neglected).

The LLL GL model was also studied numerically in both Lawrence - Doniah
model (a good approximation of the 3D GL for large number of layers) \cite%
{Sasik,Hu} and\ in 2D \cite{MC} and by a variety of nonperturbative
analytical methods. Among them the density functional \cite{Herbut}, $1/N$ %
\cite{Affleck,Moore2,Lopatin}, dislocation theory of melting \cite{Maniv}
and others \cite{Andreev}.

As we show in this paper, the BP liquid free energy combined with the
correct two loop solid energy computed recently gives scaled melting
temperature $a_{T}^{m}=-9.5$ and in addition predicts other characteristics
of the model. The magnetization of liquid is larger than that of solid by $%
1.8\%$ irrespective of the melting temperature (the specific heat jump is
about $6\%$ and decreases slowly with temperature in $YBCO$). Brief account
of these results was published \cite{Lirapid}.

In addition to the theory of melting, we considered overcooled liquid and
calculated magnetization and specific heat curves. Since the metastable
overcooled liquid state exists all the way down to zero temperature, we can
define liquid Madelung energy. Looking at the melting process from the low
temperature side for both the liquid and the solid we find that the Madelung
energy of liquid is larger than that of the solid approximately by the
latent heat of melting. Our magnetization curves agree quite well with Monte
Carlo simulations of the LLL GL \cite{Sasik}, and almost perfectly for
specific heat in 2D by Kato and Nagaosa in ref.\cite{MC}.

In addition we consider in this paper several ''phenomenological'' issues,
some matter of significant disagreement. First is the range of applicability
of the LLL model. We find that in order to describe experimental reversible
magnetization of $YBCO$ at lower fields, higher Landau levels (HLL)
corrections should be incorporated. We therefore clarify in section V the
role of the HLL modes. Experimentally it was claimed that one can establish
the LLL scaling for fields above $3T$ \cite{Sok}. A glance at the data
however shows that in normal state (above $T_{c}$) the LLL scaling for
magnetization curves is generally very bad. Most of the HLL effects can be
taken into account by just renormalizing parameters of the LLL model.
Therefore one should use the ''effective LLL''\ in which HLL were
''integrated out''. To clarify this often salient feature we explicitly
perform this integration within a self consistent approach in section V. It
was noted by Koshelev \cite{Koshelev} and others that, to calculate
magnetization, one has to carefully account for renormalization of the free
energy since it is field dependent. Then we calculated the leading
correction the effective LLL and compared with experiments. It is found that
although the LLL contribution to magnetization is much larger that the
experimentally observed one above $T_{c}$, it is nearly cancelled by the HLL
contributions. This explains the breaking of the LLL scaling in the normal
state.

The paper is organized as follows. The model is defined and its
applicability range discussed in section II. In section III, supercooled
liquid in general repulsive system will be discussed. We provide evidence
for the scenarios outlined above using both a solvable (for a large number
of components) Ginzburg - Landau model and analyzing numerical results in
other systems. In section IV the LLL model is solved and the melting theory
of vortex lattice is presented and compared to experiments. In section V,
the HLL corrections are discussed and the magnetization curves are compared
with experiments.

Phenomenological issues are addressed in sections IIB (assumptions), IVC
(melting line, Ginzburg parameter fit for various materials), IVD
(magnetization, entropy jumps), IVE (specific heat jumps) and VD (reversible
magnetization curve), so readers not interested in theoretical details can
directly proceed to these sections.

\section{The GL model and its basic assumptions.}

\subsection{The GL model}

On the mesoscopic scale a 3D superconducting materials with not very strong
asymmetry along the $z$ axis are effectively described by the following
Ginzburg-Landau free energy functional:
\begin{equation}
F[\psi ,\psi ^{\ast },{\bf A]}=\int d^{3}x\frac{{\hbar }^{2}}{2m_{ab}}%
\left\vert {\bf D}\psi \right\vert ^{2}+\frac{{\hbar }^{2}}{2m_{c}}|\partial
_{z}\psi |^{2}-a(T)|\psi |^{2}+\frac{b^{\prime }}{2}|\psi |^{4}+\frac{\left(
{\bf B}-{\bf H}\right) ^{2}}{8\pi }  \label{GLdef}
\end{equation}%
involving the order parameter field $\psi $ and magnetic field ${\bf B}$.
The external constant magnetic field is described by the vector potential in
Landau gauge ${\bf A}_{0}=(Hy,0,0).$ The covariant derivative is defined by $%
{\bf D}\equiv {\bf \nabla }-2\pi i{\bf A/\Phi _{0},}\Phi _{0}\equiv
hc/e^{\ast }$($e^{\ast }=2e$). The microscopic thermal fluctuations are
integrated out and, as a consequence, coefficients $a$, $b^{\prime }$ and $m$
depend on temperature. Mesoscopic thermal fluctuations of the order
parameter are described by the partition function:
\begin{equation}
Z=\int {\cal D}\psi {\cal D}\psi ^{\ast }{\cal D}{\bf A}\exp \left\{ -\frac{%
F[\psi ,\psi ^{\ast },{\bf A]}}{T}\right\}  \label{Zdefinition}
\end{equation}%
Our aim is to quantitatively describe the effects of thermal fluctuations of
high $T_{c}$ cuprates of the $YBCO$ type and some other "strongly
fluctuating" 3D materials.

\subsection{Assumptions}

The use of the above GL energy hinges upon several physical assumptions.
They are listed below.

(1) {\it Continuum 3D model}

We use the anisotropic GL model despite the well established layered
structure of the high $T_{c}$ cuprates for which models of the Lawrence -
Doniah type are more appropriate. Effects of layered structure are dominant
in $BSCCO$ or $Tl$ compounds (anisotropy very large: $\gamma \equiv \sqrt{%
m_{c}/m_{ab}}>1000$) and noticeable for cuprates with anisotropy of order $%
\gamma =50$ like $LaBaCuO,$ strongly underdoped $YBCO$ (see however \cite%
{Shapiro}) or $Hg1223$. The requirement, that the 3D GL can be effectively
used, therefore limits us to optimally doped $YBCO_{7-\delta }$ (or slightly
overdoped or underdoped) for which the anisotropy parameter is not very
large $\gamma =4-8$ \cite{Schilling2}), $DyBCO$ and possibly $Hg1221$ which
has a slightly larger anisotropy. There is no such problem in recently
discovered isotropic \textquotedblright fluctuating\textquotedblright\
superconductor $(K,Ba)BiO_{3}$ \cite{Marcus}.

(2) {\it Range of validity of the mesoscopic (GL) approach}

The GL approach generally is an effective mesoscopic approach. It is
applicable when one can neglect higher order terms in the functional eq.(\ref%
{GLdef}) typically generated, when one ''integrates out''\ microscopic
degrees of freedom. The leading higher dimensional terms we neglect (as
''irrelevant'') are $|\psi |^{6}$ and higher (four) derivative terms like $|%
{\bf D}^{2}\psi |^{2}$. This naively leads to a condition that $1-t-b$ is
smaller than $1.$ Here and in what follows%
\begin{equation}
t\equiv T/T_{c};\text{ \ \ \ }b\equiv H/H_{c2}\approx B/H_{c2}.
\end{equation}%
The applicability line $1-t-b<0.2$ for $YBCO$ is plotted in Fig. 1. We also
will consider a model invariant under rotations in the $ab$ plane.
Noninvariant models sometimes can be rescaled it to $m_{a}\simeq
m_{b}=m_{ab} $ \cite{Blatter}$.$ For several physical questions those
assumptions are not valid because neglected ''irrelevant'' terms might
become ''dangerous''. For example the question of the structural phase
transition into the square lattice is clearly of this type \cite{structural}%
. It is known that even assuming $m_{a}/m_{b}=1$ in low temperature vortex
lattices in $YBCO,$ rotational symmetry is broken down to the fourfold
symmetry by the four derivative terms. However there is no significant
correction to, for example, the magnetization from those higher dimension
terms.

(3) {\it Expansion of parameters around }$T_{c}$

Generally parameters of the GL model of eq.(\ref{GLdef}) are complicated
functions of temperature which are determined by the details of the
microscopic theory. We expand the coefficient $a(T)$ near $T_{c}:$
\begin{equation}
a(T)=T_{c}[\alpha (1-t)-\alpha ^{\prime }(1-t)^{2}+...].
\end{equation}%
The second and higher terms in the expansion are omitted and therefore, when
temperature deviates significantly from $T_{c},$ one cannot expect the model
to have a good precision. We note that recently measured$\ H_{c2}(T)$ is
linear in $T$ in a wide region near $T_{c}$ in both $YBCO$ and $%
(K,Ba)BiO_{3} $ \cite{Nakagawa,Marcus}.

(4) {\it Constant nonfluctuating magnetic field}

For strongly type II superconductors like the high $T_{c}$ cuprates not very
far from $H_{c2}(T)$ (this easily covers the range of interest in this
paper, for the detailed discussion of the range of applicability beyond it
see ref.\cite{LiHLL}) magnetic field is homogeneous to a high degree due to
superposition from many vortices. Inhomogeneity is of order $1/\kappa
^{2}\sim 10^{-3}$. Since the main subject of this study is thermal
fluctuation effects of the order parameter field, one might ask whether
thermal fluctuations of the electromagnetic field should be also taken into
account. Halperin, Lubensky and Ma considered this question long time ago %
\cite{HLM}. The conclusion was that they are completely negligible for very
large $\kappa .$ Upon discovery of the high $T_{c}$ cuprates, the issue was
reconsidered \cite{Lobb} and the same result was obtained to a very high
precision. Therefore here magnetic field is treated both as constant and
nonfluctuating ($B=H$) and the last term in eq.(\ref{GLdef}) can be omitted
(to precision of order $1/\kappa ^{2}$). However when we calculate \ the
magnetization, $M=(B-H)/4\pi $ which is of order $1/\kappa ^{2}$, higher
order correction must be considered.

Recently it was claimed that the ''vortex loops''\ fluctuations are
important and even might lead to additional phase transition at field of
order $Gi\ H_{c2}$ \cite{Tesanovic}. This is of order $\ 100G$ for the
materials of interest listed in Table 2 and therefore is irrelevant for
physics discussed in this paper. Note that $Gi$ in the papers discussing the
vortex loops \cite{Sudbo} physics is assumed to be much larger. We discuss
this issue in section IVC

(5) {\it Disorder.}

Point - like disorder is always present in $YBCO$. For example magnetization
becomes irreversible. Melting line of the optimally doped or underdoped
samples bends towards lower fields \cite{Bouquet} and signs of the second
order transition appear at $12T$ \cite{Paulius}. However in some samples
like fully oxidized $YBa_{2}Cu_{3}O_{7}$\cite{Nishizaki} and $%
DyBa_{2}Cu_{3}O_{7}$ \cite{Roulin,Garfield} the disorder effects are minor
especially at temperature close to $T_{c}$. In the maximally oxidized $YBCO$ %
\cite{Nishizaki} the second order transition associated with disorder is not
seen even at highest available fields ($30T$).Certain aspects of the
disorder problem were addressed \ in the framework of GL theory \cite%
{Lidisorder}, elasticity theory \cite{Giamarchi} and phenomenological
approach based on the Lindermann criterion \cite{Mikitik}.

\subsection{Landau level modes in the quasi momentum basis}

Assuming that all the requirements are met we now divide the fluctuations
into the LLL and HLL modes. Throughout most of the paper will use the
coherence length $\xi =\sqrt{{\hbar }^{2}/\left( 2m_{ab}\alpha T_{c}\right) }
$ as a unit of length, $T_{c}$ as unit of temperature and $\frac{%
dH_{c2}(T_{c})}{dT}T_{c}=\frac{\Phi _{0}}{2\pi \xi ^{2}}$ as a unit of
magnetic field. As we mentioned above, we assume constant magnetic induction
${\bf B}={\bf b}H_{c2}$ which is slightly different from the external
magnetic field ${\bf H}={\bf h}H_{c2}$. After rescaling eq.(\ref{GLdef})by $%
x\rightarrow \xi x,y\rightarrow \xi y,z\rightarrow \frac{\xi z}{\gamma }%
,\psi ^{2}\rightarrow \frac{2\alpha T_{c}}{b^{\prime }}\psi ^{2}$ ($\gamma
\equiv \sqrt{m_{c}/m_{ab}}$) one obtains the Boltzmann factor:
\begin{equation}
f=\frac{F}{T}=\frac{1}{\omega }\int d^{3}x\left[ \frac{1}{2}|{\bf D}\psi
|^{2}+\frac{1}{2}|\partial _{z}\psi |^{2}-\left( a_{h}+\frac{b}{2}\right)
|\psi |^{2}+\frac{1}{2}|\psi |^{4}+\frac{\kappa ^{2}\left( {\bf b}-{\bf h}%
\right) ^{2}}{4}\right] ,  \label{GLs1}
\end{equation}%
where dimensionless parameter
\begin{equation}
\omega =\sqrt{2Gi}\pi ^{2}t  \label{omega}
\end{equation}%
characterizes the strength of thermal fluctuations. The commonly used
dimensionless Ginzburg number is defined by
\begin{equation}
Gi\equiv \frac{1}{2}\left( \frac{32\pi e^{2}\kappa ^{2}\xi T_{c}\gamma }{%
c^{2}h^{2}}\right) ^{2}.  \label{Gidef}
\end{equation}%
And
\begin{equation}
a_{h}=\frac{1-t-b}{2}.
\end{equation}%
defines the distance from the mean field transition line. It is convenient
to expand the order parameter field in a complete basis of noninteracting
theory: the Landau levels. In the hexagonal lattice phase the most
convenient basis is the quasi-momentum basis:
\begin{equation}
\psi (x,y,z)=\frac{1}{\sqrt{2}\left( 2\pi \right) ^{3/2}}\int_{k}%
\sum_{n=0}^{\infty }e^{-ik_{z}z}\varphi _{{\bf k}}^{n}(x,y)\psi ^{n}({\bf k,}%
k_{z}).
\end{equation}%
Here $\varphi _{{\bf k}}^{n}(x)$ is the eigenstate of the $n^{th}$ Landau
level $\varepsilon _{n}=(n+1/2)b$ with two dimensional quasi-momentum ${\bf k%
}$ with the hexagonal symmetry:
\begin{eqnarray}
\varphi _{{\bf k}}^{n} &=&\sqrt{\frac{\sqrt{\pi }}{2^{n-1}n!a_{\vartriangle }%
}}\sum\limits_{l=-\infty }^{\infty }H_{n}(y\sqrt{b}+\frac{k_{x}}{\sqrt{b}}-%
\frac{2\pi }{a_{\vartriangle }}l)  \label{quasim} \\
&&\times \exp \left\{ i\left[ \frac{\pi l(l-1)}{2}+\frac{2\pi (\sqrt{b}x-%
\frac{k_{y}}{\sqrt{b}})}{a_{\vartriangle }}l-xk_{x}\right] -\frac{1}{2}(y%
\sqrt{b}+\frac{k_{x}}{\sqrt{b}}-\frac{2\pi }{a_{\vartriangle }}%
l)^{2}\right\} .  \nonumber
\end{eqnarray}%
where $a_{\vartriangle }\equiv \sqrt{\frac{4\pi }{\sqrt{3}}}$. The function $%
\varphi _{A}\equiv \varphi _{{\bf k=0}}^{n=0}$ describes the Abrikosov
lattice solution\cite{Tinkham}. Even in the liquid state which is more
symmetric than the hexagonal lattice, we find it convenient to use this
basis.

Naively, if the magnetic field is sufficiently high, the energy gap of the
order $b$ separating the $n=0$ LLL modes from the HLL is very large it is
reasonable to keep only the LLL modes in eq.(\ref{GLs1}). The dominance of
the LLL modes for melting was discussed in ref.(\cite{Brezin}), and Pierson
and Valls in ref. \cite{Sok} and we will discuss it in more detail in
section V. In the rest of this section, we consider the LLL GL model.

\subsection{The LLL scaling}

Using the LLL condition $|{\bf D}\psi |^{2}=b|\psi |^{2}$, the free energy
simplifies:
\begin{equation}
f=\frac{1}{\omega }\int d^{3}x\left[ \frac{1}{2}|\partial _{z}\psi
|^{2}-a_{h}|\psi |^{2}+\frac{1}{2}|\psi |^{4}+\frac{\kappa ^{2}\left( {\bf b}%
-{\bf h}\right) ^{2}}{4}\right] .
\end{equation}%
There is no longer a gradient term in directions perpendicular to the field
and consequently the model possesses the LLL scaling \cite{Thouless}. After
additional rescaling $x\rightarrow x/\sqrt{b},y\rightarrow y/\sqrt{b}%
,z\rightarrow z\left( \frac{b\omega }{4\pi \sqrt{2}}\right) ^{-1/3},\psi
\rightarrow \left( \frac{b\omega }{4\pi \sqrt{2}}\right) ^{1/3}\psi $, the
dimensionless free energy takes a form:
\begin{equation}
f=\frac{1}{4\pi \sqrt{2}}\int d^{3}x\left[ \frac{1}{2}|\partial _{z}\psi
|^{2}+a_{T}|\psi |^{2}+\frac{1}{2}|\psi |^{4}+\kappa ^{2}\left( \frac{%
b\omega }{4\pi \sqrt{2}}\right) ^{-4/3}\frac{\left( {\bf b}-{\bf h}\right)
^{2}}{4}\right] .  \label{LLL2}
\end{equation}%
Minimizing it with respect to $b$ leads to magnetization $b-h$ of the order\
$\frac{1}{\kappa ^{2}}$ . This means that in the strongly type II limit ($%
\kappa >>1$) the last term is of the order $1/\kappa ^{2}$ and can be
neglected. The theory has a single dimensionless parameter, the Thouless
scaled temperature defined by:
\begin{mathletters}
\begin{equation}
a_{T}=-\left( \frac{b\omega }{2^{5/2}\pi }\right) ^{-2/3}a_{h}.  \label{athl}
\end{equation}%
The Gibbs free energy density in the newly scaled model is defined as:
\end{mathletters}
\begin{equation}
g(a_{T})=-\frac{4\pi \sqrt{2}}{V}\log \int D\psi D\psi ^{\ast }\exp \left\{
-f[\psi ]\right\} ;
\end{equation}%
which is also a function of $a_{T}$ only ($4\pi \sqrt{2}$ is the scaled
``temperature''.). The relation to the original Gibbs free energy is:%
\begin{equation}
G(T,H)=\frac{H_{c2}^{2}}{2\pi \kappa ^{2}}\left( \frac{b\omega }{2^{5/2}\pi }%
\right) ^{4/3}g(a_{T}).  \label{Gdef}
\end{equation}

\section{Overcooled liquid and the T=0 fixed point of the LLL model}

\subsection{Mean field approximation: absence of the finite temperature
pseudo critical point for the vortex liquid.}

The energy of the hexagonal solid in mean field (neglecting mesoscopic
thermal fluctuations) is \cite{Tinkham}:
\begin{equation}
g_{M}^{sol}=-\frac{a_{T}^{2}}{2\beta _{A}};\text{ \ \ \ }G_{M}^{sol}=-\frac{%
H_{c2}^{2}}{4\pi \kappa ^{2}\beta _{A}}a_{h}^{2}  \label{gmad}
\end{equation}%
where $\beta _{A}=1.1596$ and the subscript "$\acute{M}$" underlies
similarity to the Madelung energy of atomic solids. The major fluctuations
contribution to the solid\ free energy is due to the \textquotedblright
phonon\textquotedblright\ modes. In harmonic approximation it is
proportional to the fluctuation temperature $T=a_{T}^{-3/2}:$%
\begin{eqnarray}
g_{1}^{sol} &=&2.848\left\vert a_{T}\right\vert ^{1/2};\text{ \ \ \ \ }%
G_{1}^{sol}=C_{1}^{sol}T;  \label{g1sol} \\
C_{1}^{sol} &=&2.848\frac{H_{c2}B}{8\kappa ^{2}T_{c}}\sqrt{|a_{h}|}.
\nonumber
\end{eqnarray}%
At low fluctuation temperatures one can neglect the $T$ dependence of $%
a_{h}\simeq -(1-b)/2$. Solid becomes unstable at $a_{T}=-5\ $according to
the self consistent (gaussian) approximation \cite{LiSolid}.

In the (homogeneous) liquid state order parameter vanishes and the
contributions to free energy come solely from fluctuations. The gaussian
(''mean field'')\ approximation to the free energy \cite{Ruggeri} is
\begin{equation}
g=4\sqrt{\varepsilon }-4/\varepsilon ,  \label{gmean}
\end{equation}%
where the excitation energy $\varepsilon $ is given by a solution of the
cubic ''gap equation''\
\begin{equation}
\varepsilon ^{3/2}-a_{T}\sqrt{\varepsilon }-4=0.  \label{gapliq}
\end{equation}%
The liquid state becomes metastable below the melting temperature, but
unlike the solid above melting, does not loose metastability at certain
''spinodal''\ point \cite{Debenedetti}. It persists all the way down to $T=0$%
. The excitation energy of the supercooled liquid approaches zero as a power
$\varepsilon \sim 16/a_{T}^{2}$. For $a_{T}\rightarrow -\infty $, the scaled
energy eq.(\ref{gmean}) has an expansion in $1/a_{T}^{3}\propto T^{2}$ for
small fluctuation temperature $T$ (the radius of convergency of the
expansion extending to $a_{T}=-3$). Therefore the liquid despite having
energy larger than that of solid becomes (pseudo) critical \cite{Compagner}
at zero temperature. Physical quantities ''around''\ this point exhibits a
power behavior with characteristic (pseudo) critical exponents. The
metastable liquid state has a distinct Madelung energy%
\begin{equation}
G_{M}^{liq}=-\frac{H_{c2}^{2}}{8\pi \kappa ^{2}}a_{T}^{2}.
\end{equation}%
As temperature increases the difference between the solid and the liquid
becomes smaller and vanishes at melting. Generally one expects a linear
correction at small $T$:
\begin{equation}
G^{liq}=G_{M}^{liq}+C_{1}^{liq}T.  \label{gliq}
\end{equation}%
Since the expansion of the mean field free energy is in $T^{2}$: $%
C_{1}^{liq}=0$. Comparing the solid free energy eqs.(\ref{gmad},\ref{g1sol})
with eq.(\ref{gliq}), we get the melting temperature $a_{T}^{m}=-6.3$. We
therefore conclude that in this approximation the supercooled liquid state
exists down to its pseudo critical point \ at zero temperature. Moreover the
pseudo critical point might govern the behavior of the liquid phase to
temperature as high as the melting point.

\subsection{The large N approximation and the Lopatin - Kotliar model of the
Abrikosov lattice melting}

It is important to confirm the above scenario in an exactly solvable model.
The simplest model of this kind is the multicomponent GL model. The LLL GL
theory can be generalized (in several different ways) to an $N$ component
order parameter field $\psi ^{a},a=1,...,N:$

\begin{equation}
f=\frac{1}{4\pi \sqrt{2}}\int d^{3}x\left[ \frac{1}{2}|\partial _{z}\psi
^{a}|^{2}+a_{T}|\psi ^{a}|^{2}+\frac{\nu }{2N}|\psi ^{a}|^{2}|\psi ^{b}|^{2}+%
\frac{1-\nu }{2N}\psi ^{a}\psi ^{a}\psi ^{\ast b}\psi ^{\ast b}\right] .
\label{Kotliar}
\end{equation}%
The large $N$ limit of this theory can be solved in a way similar to that in
the $N$ component scalar models widely used in theory of critical phenomena %
\cite{Amit}. The simplest case $\nu =1$ has been considered in ref. \cite%
{Affleck}. It was found that the homogeneous state is stable at all
temperatures. Under assumption that the conventional Abrikosov lattice takes
over at low temperatures it supported the original conjecture by Brezin et
al. \cite{Brezin} that melting of the flux lattice is a first order phase
transition. However it was shown (by explicit numerical evaluation) in \cite%
{MooreRG} that the low temperature ground state in that model is not the
Abrikosov lattice state in which just one component has a nonzero
expectation value (similar to the one component Abrikosov lattice). The
''true'' ground state has infinite degeneracy. Different ground states at
large $N$ are markedly different from the hexagonal lattice. The case $v=2,$%
.in which the Abrikosov lattice state is a stable ground state, was first
introduced in \cite{Lopatin} (in what follows it will be referred to as the
LK model). Eq.(\ref{Kotliar}) is a slight generalization including both
models studied in ref. \cite{Affleck,Lopatin}. We find that in fact all
models with $\nu \geq 2$ possess such a stable lattice state.

A straightforward method to develop the $1/N$ expansion with the last
component of $\psi ^{N}$ having the expectation value $\propto \varphi
_{A}\equiv \varphi _{{\bf k=0}}^{n=0},$ describing the hexagonal lattice
(see see eq.(\ref{quasim})), is to shift this field $\psi
^{N}(x,y,z)\rightarrow \psi ^{N}(x,y,z)+\sqrt{N}c\varphi _{A}(x,y),$ where $%
c $ is a (real) constant. Then one introduces Hubbard - Stratonovich (HS)
fields $\rho ,\chi $ \cite{Lopatin} via free energy:

\begin{eqnarray}
f[\psi ^{a},\rho ,\chi ] &=&\frac{1}{4\pi \sqrt{2}}\left\langle \frac{1}{2}%
|\partial _{z}\psi ^{a}|^{2}+(\nu \rho +a_{T})|\psi ^{a}|^{2}+\nu
c^{2}|\varphi _{A}|^{2}|\psi ^{a}|^{2}+\frac{1-\nu }{2}\left[ \left(
c^{2}\varphi _{A}^{2}+\chi )\psi ^{\ast b}\psi ^{\ast b}+cc\right) \right]
\right\rangle _{x}  \nonumber \\
&&-\frac{N}{4\pi \sqrt{2}}\left\langle \frac{\nu }{2}\rho ^{2}+\frac{1-\nu }{%
2}|\chi |^{2}\right\rangle _{x}+N\ f_{nf}+...
\end{eqnarray}%
Here the ''nonfluctuating part'' is the Abrikosov free energy density%
\begin{equation}
f_{nf}=\frac{1}{4\pi \sqrt{2}}\left[ a_{T}c^{2}+\frac{\beta _{A}}{2}c^{4}%
\right] .
\end{equation}%
We omitted several cubic terms which do not influence the leading order in $%
1/N$. Integrating over the fluctuating the $\psi ^{a}$ fields one obtains
the effective scaled Gibbs energy density (the calculation is very similar
to that in \cite{LiSolid}, where technical details can be found):
\begin{equation}
\frac{g_{eff}}{N}=a_{T}c^{2}+\frac{\beta _{A}}{2}c^{4}-\left\langle \frac{%
\nu }{2}\rho ^{2}+\frac{1-\nu }{2}|\chi |^{2}\right\rangle
_{x}+2\left\langle \sqrt{\epsilon _{O}({\bf k})}+\sqrt{\epsilon _{A}({\bf k})%
}\right\rangle _{{\bf k}}.
\end{equation}%
The spectrum has two branches:%
\begin{equation}
\varepsilon _{O,A}({\bf k})=a_{T}+v\left( \beta _{k}c^{2}+\rho _{k})\pm
\left| (1-v)\left( c^{2}\gamma _{k}+\chi _{k}\right) \right| \right) .
\end{equation}%
To have a stable {\it perturbative} Abrikosov solution, the spectrum should
be positive definite for $\rho _{k}=\chi _{k}=0$. Thus we demand $-\nu
/2+(\nu -1)\geq 0$ or $\nu \geq 2,$ as stated above. Here both HS fields%
\begin{equation}
\rho _{k}=\left\langle \rho (x)|\varphi _{k}(x)|^{2}\right\rangle _{x};\text{
\ \ \ }\chi _{k}=\left\langle \chi (x)\varphi _{k}^{\ast }(x)\varphi
_{-k}^{\ast }(x)\right\rangle _{x}
\end{equation}%
and the constant $c$ should minimize free energy $g_{eff}$.

\subsection{The inhomogeneous (solid) solution}

The minimization with respect to $\rho \left( x\right) $ and $\chi \left(
x\right) $ leads to
\begin{eqnarray}
\rho \left( x\right) &=&\left\langle |\varphi _{k}(x)|^{2}\left\{ \left[
\epsilon _{O}({\bf k})\right] ^{-1/2}+\left[ \epsilon _{A}({\bf k})\right]
^{-1/2}\right\} \right\rangle _{k} \\
sign\left( 1-\nu \right) \chi \left( x\right) &=&\left\langle \varphi
_{k}(x)\varphi _{-k}(x)\frac{c^{2}\gamma _{k}+\chi _{k}}{\left| c^{2}\gamma
_{k}+\chi _{k}\right| }\left\{ \left[ \epsilon _{O}({\bf k})\right] ^{-1/2}-%
\left[ \epsilon _{A}({\bf k})\right] ^{-1/2}\right\} \right\rangle _{k},
\nonumber
\end{eqnarray}%
which, in terms of Fourier harmonics of the hexagonal lattice, takes a form:
\begin{eqnarray}
\rho _{l} &=&\left\langle \beta _{l-k}\left\{ \left[ \epsilon _{O}({\bf k})%
\right] ^{-1/2}+\left[ \epsilon _{A}({\bf k})\right] ^{-1/2}\right\}
\right\rangle _{k} \\
sign\left( 1-\nu \right) \chi _{l} &=&\left\langle \gamma _{k,l}^{\ast }%
\frac{c^{2}\gamma _{k}+\chi _{k}}{\left| c^{2}\gamma _{k}+\chi _{k}\right| }%
\left\{ \left[ \epsilon _{O}({\bf k})\right] ^{-1/2}-\left[ \epsilon _{A}(%
{\bf k})\right] ^{-1/2}\right\} \right\rangle _{k}.  \nonumber
\end{eqnarray}%
The lattice functions $\beta _{k},\gamma _{k},\gamma _{k,l}$ are defined in
appendix of \cite{LiSolid}. The only consistent solution preserving the
hexagonal symmetry is $\chi _{k}=\chi _{c}\gamma _{k},$ and the above
equation will simplify to:
\begin{equation}
sign\left( 1-\nu \right) \chi _{c}\beta _{A}=\left\langle sign\left(
c^{2}+\chi _{c}\right) \eta _{k}\left\{ \left[ \epsilon _{O}({\bf k})\right]
^{-1/2}-\left[ \epsilon _{A}({\bf k})\right] ^{-1/2}\right\} \right\rangle
_{k}.
\end{equation}%
For the LK model \cite{Lopatin}, $\nu =2,$this leads to: $\chi _{c}\beta
_{A}=\left\langle \eta _{k}\left\{ \left[ \epsilon _{A}({\bf k})\right]
^{-1/2}-\left[ \epsilon _{O}({\bf k})\right] ^{-1/2}\right\} \right\rangle
_{k}$.\ Finally the set of the minimization equations ($\chi \geq 0$) is
\begin{eqnarray}
0 &=&a_{T}+\beta _{A}c^{2}+2\left\langle \beta _{k}\left\{ \left[ \epsilon
_{O}({\bf k})\right] ^{-1/2}+\left[ \epsilon _{A}({\bf k})\right]
^{-1/2}\right\} \right\rangle _{{\bf k}}+  \nonumber \\
&&\left\langle \eta _{k}\left\{ \left[ \epsilon _{O}({\bf k})\right] ^{-1/2}-%
\left[ \epsilon _{A}({\bf k})\right] ^{-1/2}\right\} \right\rangle _{{\bf k}}
\nonumber \\
\chi _{c}\beta _{A} &=&\left\langle \eta _{k}\left\{ \left[ \epsilon _{A}(%
{\bf k})\right] ^{-1/2}-\left[ \epsilon _{O}({\bf k})\right] ^{-1/2}\right\}
\right\rangle _{k}  \label{min} \\
\rho _{l} &=&\left\langle \beta _{l-k}\left\{ \left[ \epsilon _{O}({\bf k})%
\right] ^{-1/2}+\left[ \epsilon _{A}({\bf k})\right] ^{-1/2}\right\}
\right\rangle _{k}  \nonumber
\end{eqnarray}%
and
\begin{equation}
\varepsilon _{O,A}({\bf k})=a_{T}+2\beta _{k}c^{2}+2\rho _{k}\pm \left(
c^{2}+\chi _{c}\right) \gamma _{k}.
\end{equation}%
Following formulas will be useful \ for the calculation of the free energy:

\begin{eqnarray}
\left\langle \rho ^{2}\right\rangle &=&\left\langle \beta _{l-k}\left\{
\left[ \epsilon _{O}({\bf k})\right] ^{-1/2}+\left[ \epsilon _{A}({\bf k})%
\right] ^{-1/2}\right\} \left\{ \left[ \epsilon _{O}({\bf l})\right] ^{-1/2}+%
\left[ \epsilon _{A}({\bf l})\right] ^{-1/2}\right\} \right\rangle _{k,l}
\nonumber \\
\left\langle |\chi |^{2}\right\rangle &=&\frac{1}{\beta _{A}}\left[
\left\langle \eta _{k}\left\{ \left[ \epsilon _{A}({\bf k})\right] ^{-1/2}-%
\left[ \epsilon _{O}({\bf k})\right] ^{-1/2}\right\} \right\rangle _{k}%
\right] ^{2}
\end{eqnarray}

The equations in eq.(\ref{min}) can be solved using mode expansion \cite%
{Lopatin,LiSolid}. The spectrum can be written as follows

\begin{equation}
\epsilon _{O}({\bf k})=E(k)+\Delta \eta _{k},\epsilon _{A}({\bf k}%
)=E(k)-\Delta \eta _{k},
\end{equation}%
with $E(k)$ expanded in modes
\begin{equation}
E(k)=\sum E_{n}\beta _{n}(k),
\end{equation}%
where
\begin{equation}
\beta _{k}=\sum_{n=0}^{\infty }\exp [-2\pi n/\sqrt{3}]\beta _{n}(k),\ \
\beta _{n}(k)\equiv \sum_{\left| {\bf X}\right| ^{2}=4\pi n/\sqrt{3}}\exp [i%
{\bf k\bullet X}].
\end{equation}%
The integer $n$ determines the distance of a points on reciprocal lattice
from the origin. The effective ''expansion parameter'' is $\exp [-2\pi /%
\sqrt{3}]=0.0265$ and coefficients decrease exponentially with $n$ \cite%
{LiSolid}as can be seen from Table 1.

\begin{center}
{\bf Table 1}

Coefficients of the mode expansion for the solid solution

\begin{tabular}{|c|c|c|c|c|c|}
\hline
$a_{T}$ & $g$ & $E_{1}$ & $E_{2}$ & $E_{3}$ & $\Delta $ \\
\hline
$-4.6179$ & $-3.43164$ & $0.728715$ & $-0.0022412$ & $-0.00001227$ & $0.6167$
\\
\hline
$-5$ & $-4.96636$ & $1.92669$ & $0.0717767$ & $0.00003881$ & $2.0331$ \\
\hline
$-10$ & $-34.3165$ & $6.29543$ & $0.355908$ & $0.00023872$ & $7.2718$ \\
\hline
$-20$ & $-159.826$ & $13.8477$ & $0.842385$ & $0.00058357$ & $16.3036$ \\
\hline
\end{tabular}
\end{center}

The solution disappears at $a_{T}=-4.6179$. At this point the solid is no
longer a metastable state. $\epsilon _{A}({\bf k})$ is a gapless mode and $%
\epsilon _{A}({\bf k})$ $\longrightarrow const.{\bf k}^{2}$ for ${\bf k}%
\longrightarrow 0$. For perturbative spectrum, $\epsilon _{A}({\bf k})$ $%
\longrightarrow const.{\bf k}^{4}$ for ${\bf k}\longrightarrow 0$.

\subsection{Melting in the LK model}

The energy corresponding to the solid solution of the minimization equation
eq.(\ref{min}) calculated from
\begin{equation}
\frac{g_{eff}}{N}=a_{T}c^{2}+\frac{\beta _{A}}{2}c^{4}-\left\langle \rho
^{2}-\frac{1}{2}|\chi |^{2}\right\rangle _{x}+2\left\langle \sqrt{\epsilon
_{O}({\bf k})}+\sqrt{\epsilon _{A}({\bf k})}\right\rangle _{{\bf k}}
\end{equation}%
is given in Table 1. The convergence of the mode expansion is exponential.

For liquid, we impose rotationally invariant Ansatz with $c^{2}=0,\chi =0$
and obtain the gap equation%
\begin{equation}
\rho =\frac{2}{\sqrt{a_{T}+2\rho }},
\end{equation}%
which minimizes energy
\begin{equation}
g_{liq}=-\rho ^{2}+4\sqrt{a_{T}+2\rho }.
\end{equation}%
The results for both the liquid and the solid free energy are ploted on
Fig.2. The melting point appears at $a_{T}=-5.15$.

It is well approximated in the whole region by its low temperature expansion
in powers of $|a_{T}|^{-3/2}$ (which is proportional to the ''fluctuation
temperature'' $T$ assuming that at low temperatures $a_{h}\simeq -(1-b)/2$)
\begin{eqnarray}
\frac{g^{sol}}{a_{T}^{2}} &=&c_{M}^{sol}+c_{1}^{sol}T+c_{2}^{sol}T^{2}...,T%
\equiv |a_{T}|^{-3/2}, \\
c_{M}^{sol} &=&-\frac{1}{2\beta _{A}}%
;c_{1}^{sol}=2.84835;c_{2}^{sol}=-2.54087.  \nonumber
\end{eqnarray}%
The first two terms are the same as for the one component model, while the
two loop correction is different.

Similarly the liquid energy can be expanded, but this time in powers of
square of the ''fluctuation temperature $T$
\begin{eqnarray}
\frac{g^{_{liq}}}{a_{T}^{2}}
&=&c_{M}^{liq}+c_{1}^{liq}T+c_{2}^{liq}T^{2}...,T=|a_{T}|^{-3/2}, \\
c_{M}^{liq} &=&-\frac{1}{4};c_{1,3,..}^{liq}=0;c_{2}^{liq}=6;c_{4}^{liq}=-20.
\nonumber
\end{eqnarray}%
Here the first term is the ''Madelung energy''\ of liquid at zero
fluctuation temperature. Note that, as in the mean field approximation to
the one component theory, there is no term linear in $T$ (the harmonic
approximation). This means that the specific heat vanishes at zero
temperature. Retaining just the Madelung and the harmonic term for solid we
estimate the melting temperature in the linear approximation:%
\begin{equation}
T_{m}=\frac{c_{M}^{sol}-c_{M}^{liq}}{c_{1}^{liq}-c_{1}^{sol}}  \label{Tm}
\end{equation}%
The latent heat heat in the same approximation is:%
\begin{equation}
\Delta U=c_{M}^{sol}-c_{M}^{liq}.  \label{deltaU}
\end{equation}%
Numerically this melting temperature $T_{m}=0.064$ \ corresponding to $%
a_{T}=-6.25$ and the latent heat $\Delta U=0.18$ should be compared with the
exact results: $T_{m}=0.086$ ($a_{T}=-5.15$)$,\Delta U=0.122945.$

To conclude obtains the first order melting. Supercooled liquid persists as
a metastable state all the way to zero temperature. We emphasize that this
means that the matching of the (Borel - Pade approximant to) liquid to solid
energy at $T=0$ employed in \cite{Ruggeri} to improve convergence of the
series is not only ineffective \cite{Wilkin}, but should lead to an
incorrect result. Liquid and solid energies are different in the limit of
zero fluctuation temperature.

\subsection{General hypothesis about melting of lattices made of repelling
objects}

In atomic liquids, an attractive long range force is generally present. As a
result the supercooled liquid state loses its metastability at an end point
(spinodal) \cite{Debenedetti}. Lovett argued on general grounds long time
ago \cite{Lovett} (stability analysis of approximate set of relations
between density correlators) that for certain purely repelling interactions
the spinodal point disappears (shifted to zero temperature) and is recovered
when the attractive interaction is introduced. The existence of a metastable
overcooled liquid down to zero temperature for repelling particles therefore
might be quite general. The best studied example of mutually repelling
particles is the classical one component Coulomb plasma. We assume that, as
in the vortex system at low temperature, the supercooled liquid has a
Madelung energy and moreover its free energy has a low temperature
expansion. The free and internal energies at low temperatures can be
expanded as:%
\begin{eqnarray}
F^{sol,liq} &=&C_{M}^{sol,liq}+C_{1}^{sol,liq}T_{f}+C_{2}^{sol,liq}T_{f}^{2}-%
\frac{3}{2}T_{f}\log T_{f}+...  \label{expansion} \\
U^{sol,liq} &=&C_{M}^{sol,liq}+\frac{3}{2}T_{f}-C_{2}^{sol,liq}T_{f}^{2}+...,
\nonumber
\end{eqnarray}%
where scaled temperature $T_{f}$ is inverse of to the dimensionless plasma
parameter $\Gamma =\left( \frac{4\pi n_{s}}{3}\right) ^{1/3}\frac{e^{2}}{T}$
for density $n_{s}$. Existent Monte Carlo simulations of the internal energy
in the stable and metastable region of the 3D one component Coulomb plasma %
\cite{Caillol} can be well fitted (see solid line on Fig. 3) by
\begin{equation}
C_{M}^{liq}=-0.89186,\text{ }C_{2}^{liq}=-23.89.
\end{equation}%
It underestimates the internal energy at higher temperatures. Note that this
fit is quite different from a variety of the fractional power expressions
used at higher temperatures. One of the more successful (not very far from
the melting point) liquid theories based on density functional approach is %
\cite{Rosenfeld}:%
\begin{equation}
U^{liq}=-0.9+0.388\ T_{f}^{3/5}  \label{Rosen}
\end{equation}%
(dashed line in Fig. 3). Note however that these expression cannot be
continued to $T=0,$ since finite packing parameter is assumed. Unfortunately
the coefficient $C_{1}^{liq}$ cannot be deduced from internal energy only,
while free energy is not available at large coupling.

In the solid the analytical calculation gives \cite{Dubin}
\begin{equation}
C_{M}^{sol}=-.895929,\text{ \ \ }C_{1}^{sol}=-1.8856,\text{ \ \ }%
C_{2}^{sol}=-10.84.
\end{equation}%
This was corroborated by recent simulations \cite{Baiko}. Using the linear
in $T$ approximation eq.(\ref{deltaU}) we obtain latent heat $\Delta
U=0.0041,$ which should be compared with MC simulation result \cite{Slattery}
$\Delta U=0.043$. From the measured melting temperature \cite{Dubin} $%
T_{m}=1/172$ and coefficients fitted above one deduces
\begin{equation}
C_{1}^{liq}=-2.5.  \label{C1}
\end{equation}%
One observes that even at melting the linear approximation is justified (the
$C_{2}^{liq}$ contribution to free energy account for less than $10\%$ of
the linear one). It would be very interesting to simulate the 3D Coulomb
plasma at even stronger coupling $\Gamma >200$ to verify the existence of
expansion of supercooled liquid free energy as in eq.(\ref{expansion}).We
expect that other liquids with purely repulsive interactions like the Yukawa
(screened Coulomb) studied recently in connection with ''dusty plasma'' or
colloid suspensions physics \cite{Stevens} or even hard core repulsion lead
to qualitatively similar result. An intriguing question is whether structure
function is universal in the zero temperature limit of the liqiud phase.
Since ideal liquid is (pseudo) critical, certain universal properties are
expected.

Even more closely related to the vortex system is the quantum one component
plasma - electron gas. The quantum particle is described by a fluctuating
line very analogous to a thermally fluctuating vortex line. Here we consider
2D electron gas at zero temperature%
\begin{equation}
H=-\frac{\hslash ^{2}}{2m}%
\mathop{\displaystyle\sum}%
\limits_{i}\overrightarrow{\nabla }_{i}^{2}+\frac{1}{2}%
\mathop{\displaystyle\sum}%
\limits_{i\neq j}\frac{e^{2}}{\left| r_{i}-r_{j}\right| }.
\end{equation}%
The path integral of this system is quite analogous to a system of repelling
flux lines. The quantum fluctuations in 2DEG replace the thermal
fluctuations of the vortex system. In quantum partition \ function
\begin{equation}
Z=\int D\Psi D\Psi ^{\ast }\exp \left[ \frac{i}{\hslash }A\left( \Psi ,\Psi
^{\ast }\right) \right]
\end{equation}%
with the fermionic (Grassmannian) field $\Psi $ replacing the bosonic field
in eq. (\ref{Zdefinition}). The (Wigner) solid solution can be well
approximated by the expansion in quantum fluctuations $T_{q}=1/\sqrt{r_{s}}%
=\hslash \frac{\left( \pi n_{s}\right) ^{1/4}}{e\sqrt{m}}$\cite{Tanatar},
\[
E^{sol}=C_{M}^{sol}+C_{1}^{sol}T_{q}+C_{2}^{sol}T_{q}^{2},
\]%
where $C_{M}^{sol}=-2.2122,C_{1}^{sol}=1.6284,C_{2}^{sol}=0.058$. For
unpolarized liquid, a very good fit in wide range of densities is \cite%
{Tanatar}:
\begin{eqnarray*}
E^{liq} &=&-\frac{8\sqrt{2}}{3\pi }+T_{q}^{2}+\frac{a_{0}(T_{q}+a_{1})}{%
T_{q}^{3}+a_{1}T_{q}^{2}+a_{2}T_{q}+a_{3}} \\
a_{0} &=&-0.3568,a_{1}=1.13,a_{2}=0.9052,a_{3}=0.4165.
\end{eqnarray*}%
The fit in the low tempeature region
\[
E^{liq}=C_{M}^{liq}+C_{1}^{liq}T_{q}+...
\]%
gives liquid Madelung energy $C_{M}^{liq}=-2.18154,$ while leading
correction is $C_{1}^{liq}=1.45266.$ The transition to Wigner crystal occurs
at $T_{q}^{m}=0.174$ corresponding to $r_{s}=33.$Variational MC simulation %
\cite{Tanatar} indicates that the transition occurs at $r_{s}=37.$

\section{Borel - Pade method applied to the LLL model. Melting line,
magnetization and specific heat.}

\subsection{The BP method applied to liquid energy}

As we have seen above, within mean field the liquid branch exhibits a pseudo
critical point \cite{Compagner} at $T=0$. It is well known that in the
theory of critical phenomena one can obtain an accurate description in the
critical region by applying the Borel - Pade methods to perturbation
expansion at ''weak coupling''\cite{Baker}. In technical terms there exists
a renormalization group flow from the weak coupling fixed point towards the
strongly couple one \cite{Amit}. We therefore start with (the renormalized)
weak coupling (high temperature or non-ideal gas) expansion.

The liquid LLL (scaled) free energy is written as \cite{Ruggeri}
\begin{equation}
g_{liq}=4\varepsilon ^{1/2}[1+h\left( x\right) ].  \label{gliquid}
\end{equation}%
The function $h$ can be expanded as
\begin{equation}
h\left( x\right) =\sum c_{n}x^{n},
\end{equation}%
where the \textquotedblright small parameter\textquotedblright\ $x=\frac{1}{2%
}\varepsilon ^{-3/2}$ is defined as a solution of the Gaussian gap equation
for the excitation energy $\varepsilon ,$eq.(\ref{gapeq}).\ The coefficients
$c_{n}$ can be found in \cite{Hikami}.The consecutive approximants are
plotted on Fig.4 as dashed lines ($T1$ to $T9$, $T0$ being equivalent to the
Gaussian mean field). One clearly sees that the series are asymptotic and
can be used only at $a_{T}>-2$. One can improve on this by optimizing the
variational parameter $\varepsilon $ at each order instead of fixing it at
the first order calculation. The procedure is rather involved, see \cite%
{Liliquid}, however the optimized perturbation series is convergent with
radius of convergence about $a_{T}=-5$ (see dash dotted lines $1$ to $9$ on
Fig.4). Now we construct the BP series and compare them with the optimized
perturbation series results for $a_{T}>-5$ .

We will denote by $h_{k}\left( x\right) $ the $[k,k-1]$ BP transform \cite%
{Baker} of $h(x)$ (other BP approximants violate the correct low temperature
asymptotics).\ The BP transform is defined as%
\begin{equation}
h_{k}=\int_{0}^{\infty }\widetilde{h}_{k}\left( xt\right) \exp \left(
-t\right) dt  \label{h}
\end{equation}%
where $\widetilde{h}_{k}$ is the $[k,k-1]$ Pade transform of $%
\sum_{n=1}^{2k-1}\frac{c_{n}x^{n}}{n!}$, namely a rational function $\frac{%
\sum_{i=1}^{k}a_{i}x^{i}}{\sum_{i=1}^{k-1}b_{i}x^{i}}$ with the same
expansion at small $x$ as the original function.

\ For $k=4$ and $k=5$, the liquid energy converges to required precision ($%
0.1\%$), see Fig. 4. On this Figure only $k=3$ and $5$ are shown since $k=4$
practically coincides with the latter. In what follows we will use $h_{5}$
as the best available approximation of the liquid branch. The liquid energy
completely agrees with the optimized Gaussian expansion results \cite{LiPRL}
till its radius of convergence at $a_{T}=-5$. We therefore conclude that $%
k=5 $ is quite sufficient for our purposes.

Since the metastable liquid state exists at all temperatures one can
consider the $T=0$ limit. One finds:
\begin{equation}
\frac{g^{_{liq}}(a_{T})}{g^{sol}(a_{T})}\longrightarrow 0.964
\end{equation}%
for $a_{T}\rightarrow -\infty $. For $g^{sol}(a_{T})$, the leading term in
this limit is $-\frac{a_{T}^{2}}{2\beta _{A}}$, which is the Madelung energy
of the solid. The leading term \ for $g^{liq}(a_{T})$ is $-0.964\frac{%
a_{T}^{2}}{2\beta _{A}}$. Usually the Madelung energy for the solid phase of
the point particle system is realized by minimizing the potential energy of
the system (the minimum is often obtained by taking the hexagonal lattice
for the repulsive system in 2D). In this vortex system, we can have the
supercooled liquid down to $a_{T}\rightarrow -\infty $ ($T\rightarrow 0$).
The leading term for the overcooled liquid energy, or the Madelung energy of
the liquid is therefore equal to $-0.964\frac{a_{T}^{2}}{2\beta _{A}}$,
which is slightly larger than the Madelung energy of the solid. This limit,
the "ideal liquid", however cannot be thought as a minimization of a
potential energy.

\subsection{Melting line. Comparison with Monte Carlo simulations and
Lindemann criterion.}

The solid energy to two loops is \cite{Rosenstein,LiSolid}:
\begin{equation}
g_{sol}=-\frac{a_{T}^{2}}{2\beta _{A}}+2.848\left| a_{T}\right| ^{1/2}+\frac{%
2.4}{a_{T}}.  \label{gsolid}
\end{equation}%
where $\beta _{A}=1.1596$. On Fig. 1 of ref. \cite{Lirapid} we plot the
energies of solid and liquid. They are very close near melting (see the
difference on inset of this figure). We find that the melting point is:%
\begin{equation}
a_{T}^{m}=-9.5.
\end{equation}%
The available 3D Monte Carlo simulations \cite{Sasik} unfortunately are not
precise enough to provide an accurate melting point since the LLL scaling is
violated and one gets values of $a_{T}^{m}=-14.5,-13.2,-10.9$ at magnetic
fields $1,2,5T$ respectively. We found also that the theoretical
magnetization calculated by using parameters given by ref.\cite{Sasik} is in
a very good agreement with the Monte Carlo simulation result of ref.\cite%
{Sasik}. However the determination of melting temperature needs higher
precision, and the sample size ($\sim 100$ vortices) used in ref.\cite{Sasik}
may be not large enough to give an accurate determination of the melting
temperature (due to boundary effects, LLL scaling will be violated too). The
situation in 2D is better since the sample size is much larger. We performed
similar calculation for the 2D LLL GL liquid free energy, combined it with
the earlier solid energy calculation \cite{Rosenstein,LiSolid}
\begin{equation}
g_{sol}=-\frac{a_{T}^{2}}{2\beta _{A}}+2\log \frac{\left| a_{T}\right| }{%
4\pi ^{2}}-\frac{19.9}{a_{T}^{2}}-2.92.  \label{perbresl}
\end{equation}%
and find that the melting point $a_{T}^{m}=-13.2$. It is \ in good agreement
with MC simulations \cite{MC}.

Phenomenologically melting line can be located using Lindemann criterion or
its more refined version using Debye - Waller factor. The more refined
criterion is required since vortices are not point like. It was found
numerically for Yukawa gas \cite{Stevens} that the Debye - Waller factor $%
e^{-2W}$ (ratio of the structure function at the second Bragg peak at
melting to its value at $T=0$) is about $60\%$ at the melting point. Using
methods of \cite{LiCorr}, one obtains for the 3D LLL GL model
\begin{equation}
e^{-2W}=0.59.
\end{equation}

\subsection{Fitting of the melting line. Values of the Ginzburg numbers of
various superconductors}

In this subsection we use the above results to fit experimental melting line
of several ''fluctuating''\ superconductors. As an example on Fig.2 of ref. %
\cite{Lirapid} we presented the fitting of the melting line of fully
oxidized $YBa_{2}Cu_{3}O_{7}$ \cite{Nishizaki}. Melting lines of two
different samples of the optimally doped untwinned \cite%
{Schilling,Schilling2} near $T_{c}$ ($YBa_{2}Cu_{3}O_{7-\delta }$), $%
DyBa_{2}Cu_{3}O_{7}$ \cite{Roulin} and $(K,Ba)BiO_{3}$ \cite{Marcus} are
also fitted extremely well. The results of fitting are given in Table 2 (To
remind our convention, $H_{c2}$ is defined as $T_{c}\frac{dH_{c2}(T)}{dT}%
|_{T=T_{c}}$ rather than (often inaccessible) $H_{c2}(T=0))$.

\begin{center}
{\bf Table 2}

Parameters of high $T_{c}$ superconductors deduced from the melting line.

\begin{tabular}{|c|c|c|c|c|c|c|}
\hline
material & $T_{c}$ & $H_{c2}$ & $Gi$ & $\kappa $ & $\gamma $ & reference \\
\hline
$YBCO_{7-\delta }$ & $93.1$ & $167.5$ & $1.9\times 10^{-4}$ & $48.5$ & $7.76$
& \cite{Schilling} \\ \hline
& $92.6$ & $190$ & $2\times 10^{-4}$ & $50$ & $8.3$ & \cite{Schilling2} \\
\hline
$YBCO_{7}$ & $88.2$ & $175.9$ & $7.0\times 10^{-5}$ & $50$ & $4$ & \cite%
{Nishizaki} \\ \hline
$DyBCO_{6.7}$ & $90.1$ & $163$ & $3.2\times 10^{-5}$ & $33.77$ & $5.3$ & %
\cite{Roulin} \\ \hline
$(K,Ba)BiO_{3}$ & $31$ & $26$ & $5.3\times 10^{-5}$ & $107$ & $1$ & \cite%
{Marcus} \\ \hline
\end{tabular}
\end{center}

Our value for the Ginzburg number of $YBCO$ and $DyBCO$ estimated here are
generally lower than the ones commonly believed in the literature. Often
quoted value for $YBCO$ is of order $Gi=0.01$ (see page $1134$ of commonly
used ref.(\cite{Blatter}) ). The direct calculation from eq.(\ref{Gidef})
gives $Gi=0.003$ for $\lambda =1400$ $A$, $\xi =15$ $A$, and $\gamma $ $=7$ (%
$\kappa =93.3$). Note however that these values are estimated from
measurements at very low temperature. Our values of $\lambda $ and $\xi $
are fitted to the vortex physics experiments near $T_{c}$ and extrapolating
using (admittedly questionable) two liquid model to $T=0$ give $\lambda =931$
$A$, $\xi =18.7$ $A.$ Our values of $\frac{dH_{c2}(T)}{dT}$ near $T_{c}$ are
consistent with recent measurement \cite{Nakagawa} (about $2$) and smaller
than earlier ones. There is no consensus on values of $\kappa $ measured
using the microwave technique at very low temperatures, however they are
also generally smaller than $100$ (smaller than $70$ at $T=0$ and decreasing
with temperature according to ref. \cite{Riseman} and from $50$ to $60$ in
ref. \cite{Welp2}). This explains the difference of order of magnitude in $%
Gi $ between the often used values and our fitting results (small $\kappa $
will lead a small $Gi$ as $Gi\propto \kappa ^{4}\xi ^{2}T_{c}^{2}\gamma ^{2}$
). We emphasize that the actual small parameter in the theory is not $Gi$
but rather $\omega =\sqrt{2Gi}\pi ^{2}$(see eq.(\ref{GLs1})). Even for
Ginzburg number as small as $2\times 10^{-4}$ this quantity is $0.2$. As a
result the effect of thermal fluctuations is important on a significant
portion of the phase diagram.

Recently it was found that thermal fluctuation are quite significant even in
a low $T_{c}$ material $(K,Ba)BiO_{3}$. This is despite its lower critical
temperature and very small anisotropy (and thereby very small Ginzburg
number $5.3\times 10^{-5}$). Since this material is not a ''strange metal''
nor d - wave superconductor, its $H_{c2}$ is directly accessible and there
is no problem with direct estimate of $Gi$. $\omega =0.1$ for $(K,Ba)BiO_{3}
$ is not much smaller than that of $YBCO$. There is therefore no surprise
(contrary to a statement in ref.(\cite{Marcus}) that fluctuation effects are
still experimentally observable in $(K,Ba)BiO_{3}$. In order to be able
safely ignore thermal fluctuations the fluctuation parameter $\omega $
should be of order $0.01$ in which case $Gi$ should be smaller than $5\times
10^{-7}$. These are the cases of most low $T_{c}$ materials.

\subsection{Magnetization jump at melting}

The scaled magnetization (of liquid or solid) is defined by:
\begin{equation}
m\left( a_{T}\right) =-\frac{d}{da_{T}}g\left( a_{T}\right) ,
\label{mscaled}
\end{equation}%
while the LLL contribution to the magnetization is
\begin{equation}
M_{LLL}=\frac{H_{c2}}{4\pi \kappa ^{2}}\frac{a_{h}}{a_{T}}m\left(
a_{T}\right) .  \label{magLLL}
\end{equation}%
Using expressions eqs.(\ref{gsolid}) for solid and eqs.
(\ref{gliquid},\ref{h})
for liquid the magnetization jump $\Delta M$ at the melting point $%
a_{T}^{m}=-9.5$ divided by the magnetization at the melting on the solid
side is
\begin{equation}
\frac{\Delta M}{M_{s}}=\frac{\Delta m}{m_{s}}=0.018.  \label{jumpmag}
\end{equation}%
It is indeed small and is compared on Fig.2 of ref. \cite{Lirapid} (right
inset) with experimental results of fully oxidized $YBa_{2}Cu_{3}O_{7}$ \cite%
{Nishizaki} (rhombs) and optimally doped untwinned $YBa_{2}Cu_{3}O_{7-\delta
}$ \cite{Welp} (stars). The agreement is quite good. If the HLL contribution
is significant (see next section) eq.(\ref{jumpmag}) is expected to be
violated.

\subsection{Specific heat jump at melting}

In addition to the delta function like spike at melting for specific heat
experiments, the experiments also show specific heat jump. The theory allows
us to quantitatively estimate it.

The specific heat contribution due to the vortex matter is $C=-T\frac{%
\partial ^{2}}{\partial T^{2}}G(T,H)$.\ The normalized specific heat is
defined as
\[
c=\frac{C}{C_{mf}},
\]%
where $C_{mf}=\frac{H_{c2}^{2}T}{4\pi \kappa ^{2}\beta _{A}T_{c}^{2}}$ is
the mean field specific heat of solid. Substituting the definition of the
scaled free energy eq.(\ref{Gdef}) and scaled temperature eq.(\ref{athl}),
we obtain:
\begin{eqnarray}
c &=&-\frac{16\beta _{A}}{9t^{2}}\left( \frac{b\omega }{4\pi \sqrt{2}}%
\right) ^{4/3}g(a_{T})+\frac{4\beta _{A}}{3t^{2}}\left( b-1-t\right) \left(
\frac{b\omega }{4\pi \sqrt{2}}\right) ^{2/3}g^{^{\prime }}(a_{T})  \nonumber
\\
&&-\frac{\beta _{A}}{9t^{2}}\left( 2-2b+t\right) ^{2}g^{^{^{\prime \prime
}}}(a_{T})  \nonumber
\end{eqnarray}%
Using our expressions for energy of liquid and solid we obtain the following
specific heat jump at melting:
\begin{equation}
\Delta c=0.0075\left( \frac{2-2b+t}{t}\right) ^{2}-0.20Gi^{1/3}\left(
b-1-t\right) \left( \frac{b}{t^{2}}\right) ^{2/3}.  \label{spjump}
\end{equation}%
Using the parameters of $YBCO_{7-\delta }$ obtained by fitting the melting
line, Table 2, we compare eq.(\ref{spjump}) with the experimental result of
ref. \cite{Schilling} in Fig.2 of ref. \cite{Lirapid} (right inset). Note
that error bars are very large and also that disorder might be important %
\cite{Lidisorder}, so that the agreement of the theoretical \ and
experimental result of specific jump is not good as that of magnetization
jump.

\section{Higher Landau Levels contributions. Effective LLL model.}

\subsection{Where is the LLL approximation really valid?}

Contributions of HLL are important phenomenologically in two sections of the
phase diagram. The first is at temperature above the mean field critical
temperature $T_{c}(H)$ inside the liquid phase.\ The second is far below the
melting point deep inside the solid phase.

Naively in the solid phase, when \textquotedblright distance from the mean
field transition line\textquotedblright\ is smaller than the
\textquotedblright inter Landau level gap\textquotedblright , $1-t-b<2b,$
one expects that higher Landau harmonics can be neglected. More careful
examination shows that a weaker condition $1-t-b<12b$ should be used for a
validity test of the LLL approximation \cite{LiHLL} to calculate the mean
field contributions in vortex solid. Additional factor $6$ comes from the
hexagonal symmetry of the lattice since contributions of higher Landau
levels (HLL), \ first to fifth HLL do not appear in perturbative
calculation. In the liquid state the question has been studied by Lawrie %
\cite{Lawrie} using the Hartree - Fock (gaussian) approximation. The result
was that the region of validity is limited, but quite wide, see Fig.1.

In this section we will incorporate the leading HLL correction using
Gaussian approximation and then compare the theoretical results with
experimental magnetization curves.

\subsection{Gaussian Approximation in the liquid phase}

The free energy density beyond the LLL approximation is:
\begin{equation}
G=-\frac{\omega H_{c2}^{2}}{2\pi \kappa ^{2}vol}\log \int {\cal D}\psi {\cal %
D}\overline{\psi }\exp \left( -\frac{1}{\omega }\int d^{3}x\frac{1}{2}%
|\partial _{z}\psi |^{2}-a_{h}|\psi |^{2}+\frac{1}{2}|\psi |^{4}\right) ,
\end{equation}%
where $vol$ denotes volume. In the framework of the Gaussian (Hartree -
Fock) approximation free energy\ is divided into an optimized quadratic part
$K$, and a \textquotedblright small\textquotedblright\ part $V.$ Then $K$ is
chosen in such a way that the gaussian energy is minimal. The gaussian
energy is a rigorous lower bound on energy. Due to translational symmetry of
the vortex liquid, an arbitrary $U(1)$ symmetric quadratic part $K\ $has
only one variational parameter $\varepsilon :$
\begin{equation}
K=\frac{1}{\omega }\int d^{3}x\left( \frac{1}{2}\left( |{\bf D}\psi
|^{2}-b|\psi |^{2}\right) +\frac{1}{2}|\partial _{z}\psi |^{2}+\varepsilon
|\psi |^{2}\right) .
\end{equation}%
The small perturbation is therefore:
\begin{equation}
V=\frac{1}{\omega }\int d^{3}x\left[ \left( -a_{h}-\varepsilon \right) |\psi
|^{2}+\frac{1}{2}|\psi |^{4}\right] .
\end{equation}%
The Gaussian energy consists of two parts. The first is the "Trace log"
term:
\begin{equation}
-\frac{\omega H_{c2}^{2}}{2\pi \kappa ^{2}vol}\log \left[ \int {\cal D}\psi
\exp (-K)\right] =\frac{\omega H_{c2}^{2}}{2\pi \kappa ^{2}}\frac{b}{\sqrt{2}%
\pi }\sum_{n=0}^{\infty }\sqrt{nb+\varepsilon },
\end{equation}%
The second is proportional to the expectation value of $v$ in a solvable
model defined by $K$

\begin{eqnarray}
\frac{\omega H_{c2}^{2}}{2\pi \kappa ^{2}}\left\langle V\right\rangle &=&%
\frac{\omega H_{c2}^{2}}{2\pi \kappa ^{2}}{\Huge [}\left( -a_{h}-\varepsilon
\right) \frac{b}{2\sqrt{2}\pi }\sum_{n=0}^{\infty }\frac{1}{\sqrt{%
nb+\varepsilon }}  \nonumber \\
&&+\omega \left( \frac{b}{2\sqrt{2}\pi }\sum_{n=0}^{\infty }\frac{1}{\sqrt{%
nb+\varepsilon }}\right) ^{2}{\Huge ]}.
\end{eqnarray}%
Both are divergent in the ultraviolet in a sense that at large $n$ the sums
diverge. Introducing an UV momentum cutoff which effectively limits the
number of Landau levels to $N_{f}=\frac{\Lambda }{b}-1$, the Trlog term
diverges as:
\begin{equation}
\frac{1}{\sqrt{2}\pi }b\sum_{n=0}^{\infty }\sqrt{nb+\varepsilon }=\frac{1}{%
\sqrt{2}\pi }\left[ \frac{2}{3}\Lambda ^{3/2}+\left( \varepsilon -\frac{b}{2}%
\right) \Lambda ^{1/2}\right] +u(\varepsilon ,b)  \label{regu}
\end{equation}%
with the last term, function $u$ being finite. The ''bubble''\ integral
diverges logarithmically:
\begin{equation}
\frac{b}{2\sqrt{2}\pi }\sum_{n=0}^{\infty }\frac{1}{\sqrt{nb+\varepsilon }}=%
\frac{1}{\sqrt{2}\pi }\Lambda ^{1/2}+u^{\prime },
\end{equation}%
where $u^{\prime }\equiv \frac{\partial }{\partial \varepsilon }%
u(\varepsilon ,b)$. Substituting eq.(\ref{regu}) \ into the gaussian energy
one obtains $\ $(in units of $\frac{\omega H_{c2}^{2}}{2\pi \kappa ^{2}}$):
\begin{eqnarray}
g_{Gauss} &=&\frac{1}{\sqrt{2}\pi }\frac{2}{3}\Lambda ^{3/2}+\omega (\frac{1%
}{\sqrt{2}\pi }\Lambda ^{1/2})^{2}+\left( -a_{h}-\frac{b}{2}\right) \frac{1}{%
\sqrt{2}\pi }\Lambda ^{1/2}-a_{h}u^{\prime } \\
&&+2\omega \frac{1}{\sqrt{2}\pi }\Lambda ^{1/2}u^{\prime }-\varepsilon
u^{\prime }+\omega (u^{\prime })^{2}+u.  \nonumber
\end{eqnarray}%
The first term does not depend on parameters of the system and can be
ignored (the renormalization of the reference energy density), while the
second is $\omega $ dependent and indicates that $T_{c}$ present inside $%
a_{h}$ is renormalized. Defining $a_{h}=a_{h}^{r}+2\omega \frac{1}{\sqrt{2}%
\pi }\Lambda ^{1/2}$, the above energy becomes:
\begin{eqnarray}
g_{Gauss} &=&-\omega (\frac{1}{\sqrt{2}\pi }\Lambda ^{1/2})^{2}+\left(
-a_{h}^{r}-\frac{b}{2}\right) \frac{1}{\sqrt{2}\pi }\Lambda
^{1/2}-a_{h}^{r}u^{\prime }  \label{reneq} \\
&&-\varepsilon u^{\prime }+\omega (u^{\prime })^{2}+u.  \nonumber
\end{eqnarray}%
Thus the temperature $T_{c}$ and vacuum energy will be renormalized. The
first two terms in free energy are divergent and linear in\ fluctuation
temperature $\omega $, they will not contribute to any physical quantities
like magnetization and specific heat. Minimizing the energy eq.(\ref{reneq}%
), we get\ the gap equation:
\begin{equation}
\varepsilon =-a_{h}^{r}+2\omega \ u^{\prime }  \label{gapeq}
\end{equation}%
Superscript ''r''\ will be dropped later on. The function $u(\varepsilon ,b)$
can be written in the following form
\begin{equation}
u(\varepsilon ,b)=\frac{1}{\sqrt{2}\pi }b^{3/2}v\left( \frac{\varepsilon }{b}%
\right) ,
\end{equation}%
where
\begin{equation}
v\left( x\right) =\sum_{n=0}^{\infty }\left[ \sqrt{n+x}-\frac{2}{3}(x+n+%
\frac{1}{2})^{\frac{3}{2}}+\frac{2}{3}(x+n-\frac{1}{2})^{\frac{3}{2}}\right]
-\frac{2}{3}(x-\frac{1}{2})^{\frac{3}{2}}.
\end{equation}%
For the LLL model in the Gaussian approximation, $v\left( x\right) =\sqrt{x}$%
. In the ''Prange''\ limit \cite{Prange} $Gi\rightarrow 0$, the free energy
is
\begin{equation}
\frac{\omega H_{c2}^{2}}{2\pi \kappa ^{2}}\frac{1}{\sqrt{2}\pi }%
b^{3/2}v\left( -\frac{a_{h}}{b}\right) .
\end{equation}

\subsection{Integration of the HLL modes and the effective LLL model}

A method for treating HLL modes is integrating them and obtaining an
effective LLL model. The (effective) LLL model is applicable in a
surprisingly wide range of fields and temperatures determined by the
condition that the relevant excitation energy $\varepsilon $ is much smaller
than the gap between Landau levels $b$. Within the mean field approximation
in the liquid $\varepsilon $ is a solution of the gap equation of eq.(\ref%
{gapeq}). \ For the LLL dominance region, we take a conservative condition $%
\varepsilon /\varepsilon _{c}=1/20$. One observes that, apart from the
fields smaller than $H_{LLL}\approx 0.1$ $T$ for $YBCO$, the experimentally
observed melting line and its neighborhood are well within the range of
applicability of this approximation as shown in Fig.1.

The effective LLL energy (we will use unit of energy density $\frac{%
H_{c2}^{2}}{2\pi \kappa ^{2}}$in this subsection) functional is defined by:
\begin{equation}
g_{eff}[\psi _{0}]=-\frac{\omega }{vol}\;\log \int \prod_{i=1}^{\infty }%
{\cal D}\psi _{i}{\cal D}\psi _{i}^{\ast }\exp \left\{ -f[\psi _{0},\psi
_{0}^{\ast },\psi _{i},\psi _{i}^{\ast },]\right\} ,
\end{equation}%
where $\psi _{0}$ is the LLL $N=0$ component field and the rest are denoted
by $\psi _{i}$. Expanding the\ functional up to the fourth order in $\psi
_{0}$ \ and to the second order in $\partial _{z}$ one obtains:
\begin{eqnarray}
g_{eff}[\psi _{0}] &=&\Delta g+\frac{\Delta t}{2}|\psi _{0}|^{2}+\omega \
f_{LLL}[\psi _{0}]. \\
f_{LLL}[\psi _{0}] &=&\frac{1}{\omega }\left[ \frac{1}{2}|\partial _{z}\psi
_{HLL}|^{2}-a_{h}|\psi _{HLL}|^{2}+\frac{1}{2}|\psi _{HLL}|^{4}\right] .
\nonumber
\end{eqnarray}%
The direct (no $\psi _{0}$ dependence) renormalization of energy is:
\begin{equation}
\Delta g=-\frac{\omega }{vol}\;\log \int \prod_{i=1}^{\infty }{\cal D}\psi
_{i}{\cal D}\psi _{i}^{\ast }\exp \left\{ -f_{HLL}[\psi _{i}]\right\} ,
\end{equation}%
where the HLL energy is
\begin{equation}
f_{HLL}=\frac{1}{\omega }\left[ \frac{1}{2}|\partial _{z}\psi
_{HLL}|^{2}-a_{h}|\psi _{HLL}|^{2}+\frac{1}{2}|\psi _{HLL}|^{4}\right] ,
\end{equation}%
where $\psi _{HLL}=\sum_{i=1}^{\infty }\psi _{i}.$ To calculate $\Delta g$,
\ we divide the $f_{HLL}$ into
\begin{equation}
K_{HLL}=\frac{1}{\omega }\left( \frac{1}{2}\left( |{\bf D}\psi |^{2}-b|\psi
|^{2}\right) +\frac{1}{2}|\partial _{z}\psi |^{2}+\varepsilon |\psi
|^{2}\right)
\end{equation}%
plus $f_{HLL}-K_{HLL}$. Taking $\varepsilon $ \ as the solution of the gap
equation of eq.(\ref{gapeq}), $\Delta g$ takes a form:%
\begin{eqnarray}
\Delta g &=&g_{Gauss}-g_{LLL}+2\left\langle |\psi _{0}|^{2}\right\rangle
\left( \left\langle |\psi _{0}|^{2}\right\rangle -\left\langle |\psi
|^{2}\right\rangle \right) ; \\
g_{LLL} &=&-\frac{\omega }{vol}\;\log \int {\cal D}\psi _{0}{\cal D}\psi
_{0}^{\ast }\exp \left\{ -f_{LLL}[\psi _{0}]\right\}  \nonumber
\end{eqnarray}%
Here $g_{Gauss}$ is the effective free energy of the full GL obtained in the
first subsection of the current section, eq.(\ref{reneq}), $\left\langle
|\psi |^{2}\right\rangle $ is likewise the expectation value of $|\psi |^{2}$
in the full GL. The quantity $g_{LLL}$ is the effective free energy
calculated with variational parameter $\varepsilon $\ and $\left\langle
|\psi _{0}|^{2}\right\rangle $ is the expectation value in the LLL GL. The
consistency (or matching) requirement is:
\begin{equation}
g_{eff}=-\frac{\omega }{vol}\;\log \int {\cal D}\psi _{0}{\cal D}\overline{%
\psi }_{0}\exp \left\{ -\frac{1}{\omega }g_{eff}[\psi _{0}]\right\} .
\end{equation}%
This condition determines the value of $\Delta t$:%
\begin{eqnarray}
\Delta t &=&4\left( \left\langle |\psi |^{2}\right\rangle -\left\langle
|\psi _{0}|^{2}\right\rangle \right) =4\omega \left[ u^{\prime }(\varepsilon
,b)\right] -4\left\langle |\psi _{0}|^{2}\right\rangle \\
&=&4\omega \frac{1}{\sqrt{2}\pi }b^{1/2}\left[ v^{^{\prime }}\left( \frac{%
\varepsilon }{b}\right) -\frac{1}{2}\sqrt{\frac{b}{\varepsilon }}\right] .
\nonumber
\end{eqnarray}%
For $YBCO$, the correction $\Delta t$ is small. The effective LLL GL
approach achieves a simplification by starting from the LLL effective model
with $T_{c}$ and other parameters renormalized to account for the
contribution of the HLL modes. This is what we assumed in sections III and
IV. In particular, this approach is very precise if we calculate the
properties along the melting line. For example, the magnetization jump is
mostly due to the fluctuation of the LLL modes, the background effective
energy $\Delta g$ will not contribute anything since it is the same on both
sides of the melting line.

\subsection{The HLL contribution to Magnetization}

Generally when $\kappa $ is quite large and magnetization can be
approximated by
\begin{equation}
M=-\frac{\partial }{\partial H}G(T,H).  \label{magful}
\end{equation}%
The HLL correction will be calculated as follows. We numerically solve the
gap equation (\ref{gapeq}) from which $G(T,H)$ can be obtained. Then eq.(\ref%
{magful}) is used to calculate the magnetization of the full GL model in
Gaussian approximation. The HLL correction is thus the magnetization of the
full GL model in gaussian approximation minus the magnetization of the LLL
contribution in gaussian approximation. We compare the experiments using
following approximation. While the corrections due to HLL are calculated in
gaussian approximation, the LLL contribution will be calculated
nonperturbatively. The comparison of the theoretical predictions with the
experiments for fully oxidized $YBa_{2}Cu_{3}O_{7}$\cite{Nishizaki}, is
shown on Fig. 3 of ref.\cite{Lirapid}. We used the experimental asymmetry
value $\gamma =4$ and values of $T_{c},$ $H_{c2}$ and $Gi$ from the fitting
of the melting curve (see Table 2). The agreement is fair at intermediate
magnetic fields, while at low magnetic fields is not good. It is expected
that agreement is improved at higher fields. It is not clear whether
magnetization (in contrast to magnetization jump at melting) will be
strongly influenced by disorder, so at this time it is not possible to
consider optimal doped $YBCO$ magnetization curved more quantitatively.

We comment that \ the theory of the full GL model (higher Landau levels
included) beyond Gaussian approximation is required at low magnetic fields.
Indeed experimentally it is often claimed that one can establish the LLL
scaling for fields above $3$ T for $YBCO$ (see, for example, ref. \cite{Sok}%
) as at low magnetic fields, the HLL contribution will be significant.

\section{Summary}

The problem of calculating the fluctuations effects in the framework of the
Ginzburg - Landau approach to vortex matter in type II superconductors is
sufficiently precisely solved in the LLL approximation to allow quantitative
description of the melting transition. We provided an evidence that
metastable homogeneous state (the supercooled liquid state) exists down to
zero fluctuation temperature by solving the large $N$ Ginzburg - Landau
model. Based on this understanding the supercooled liquid state is
approached using methods of physics of critical phenomena (the Borel - Pade
resummation technique). Applicability of the effective lowest Landau level
model was subsequently discussed and corrections due to higher levels is
calculated.

The theory is then applied to quantitatively describe a great variety of
experiments (confined to a region not far from $T_{c})$ including melting
curves of $YBCO,DyBCO,(K,Ba)BiO_{3}$, magnetization curves, discontinuities
of various quantities at melting.

We speculate that any system of repelling objects (examples include
classical one component plasma, electron gas...) exhibits similar features.
The supercooled metastable state extends down to zero temperature. At this
limit there is a well defined Madelung energy of the "ideal liquid". This
ideal liquid is a pseudo critical point which controls the supercooled state
possibly up to the melting temperature and might have universal features.

\acknowledgments We are grateful to E.H. Brandt, X. Hu, A. Knigavko, J.-Y.
Lin, T.Z. Uen for numerous discussions, T. Nishizaki, A. Junod and M.
Naughton for providing details of experiments and Z.Tesanovic and M. Moore
for correspondance. We are especially grateful to Y. Rosenfeld for patiently
explaining to us his results on one component plasma. The work was supported
by NSC of Taiwan grant NSC\#91-2112-M-009-503 and the Mininstrey of Science
and Technology of China (G1999064602) are acknowledged.

\bigskip


{\Huge Figure captions}

{\LARGE Fig. 1}

Comparison of the experimental melting line for fully oxidized $%
YBa_{2}Cu_{3}O_{7}$ \cite{Nishizaki} with our theoretical fitting.
Applicabilty of the LLL approximation is between two lines, the solid LLL
applicablity line and the (liquid) LLL dominace line. The GL model
applicability line is also plotted.

{\LARGE Fig. 2}

Free energy of solid (line) and liquid (dashed line) of the LargeN model as
function of the fluctuation temperature $1/|a_{T}|^{3/2}$. The solid line
ends at a\ point (dot) indicating the loss of metastability.

{\LARGE Fig. 3}

Internal energy of the classical one component Coulomb plasma. The dashed
line is the fit given by eq.(\ref{Rosen}) and the solid line is the fit
given by eq.(\ref{expansion}).

\bigskip {\LARGE Fig. 4}

The BP \ approximation for the free energy. BP3 and BP5 are the free energy
results given by $h_{3}$ and $h_{5}$. The dashed line $Ti$ is the original
perturbative expansion of order $i$ in ref. \cite{Ruggeri} and the the dot
dashed line $i$ is the optimized expansion of order $i$.

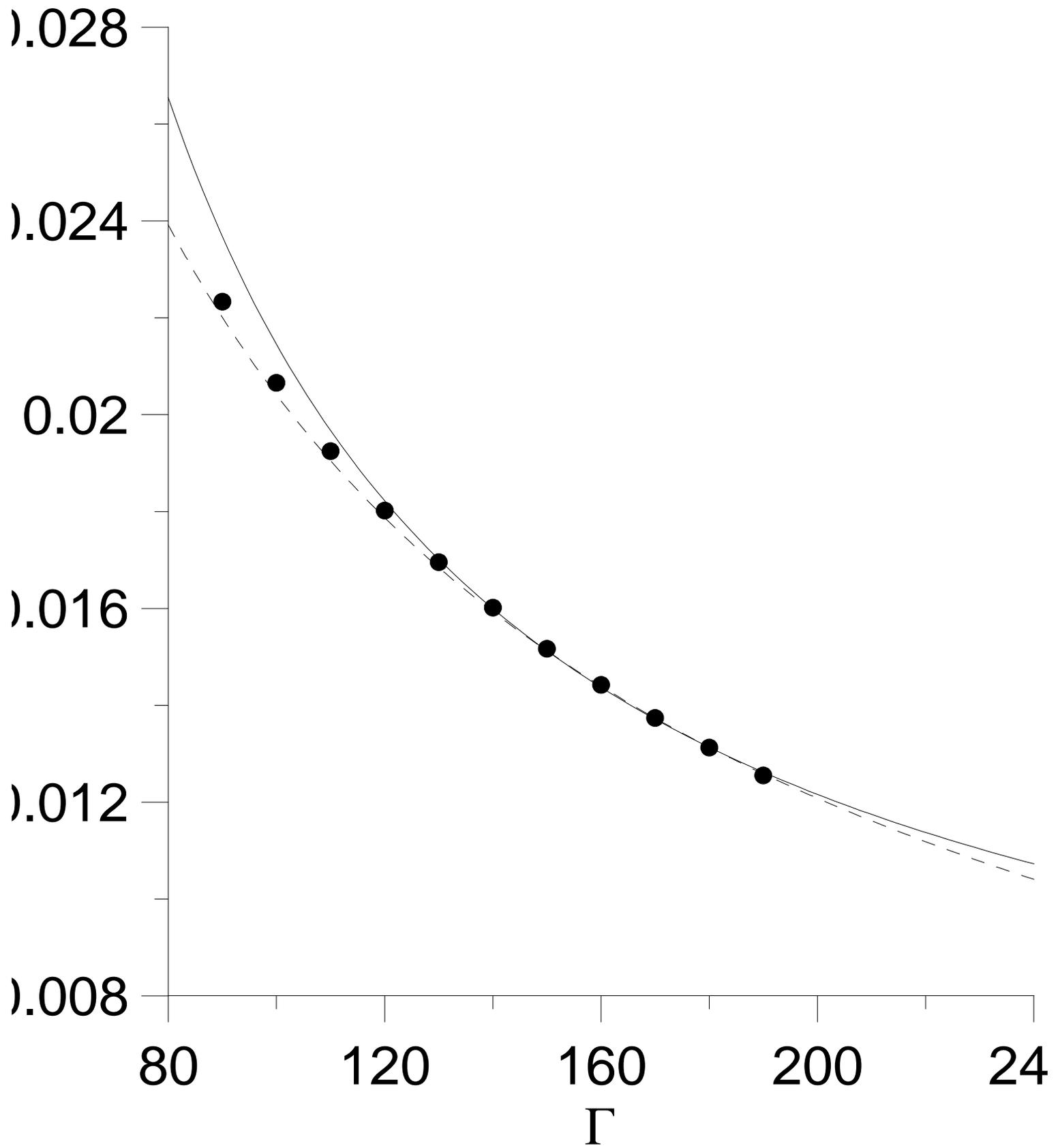
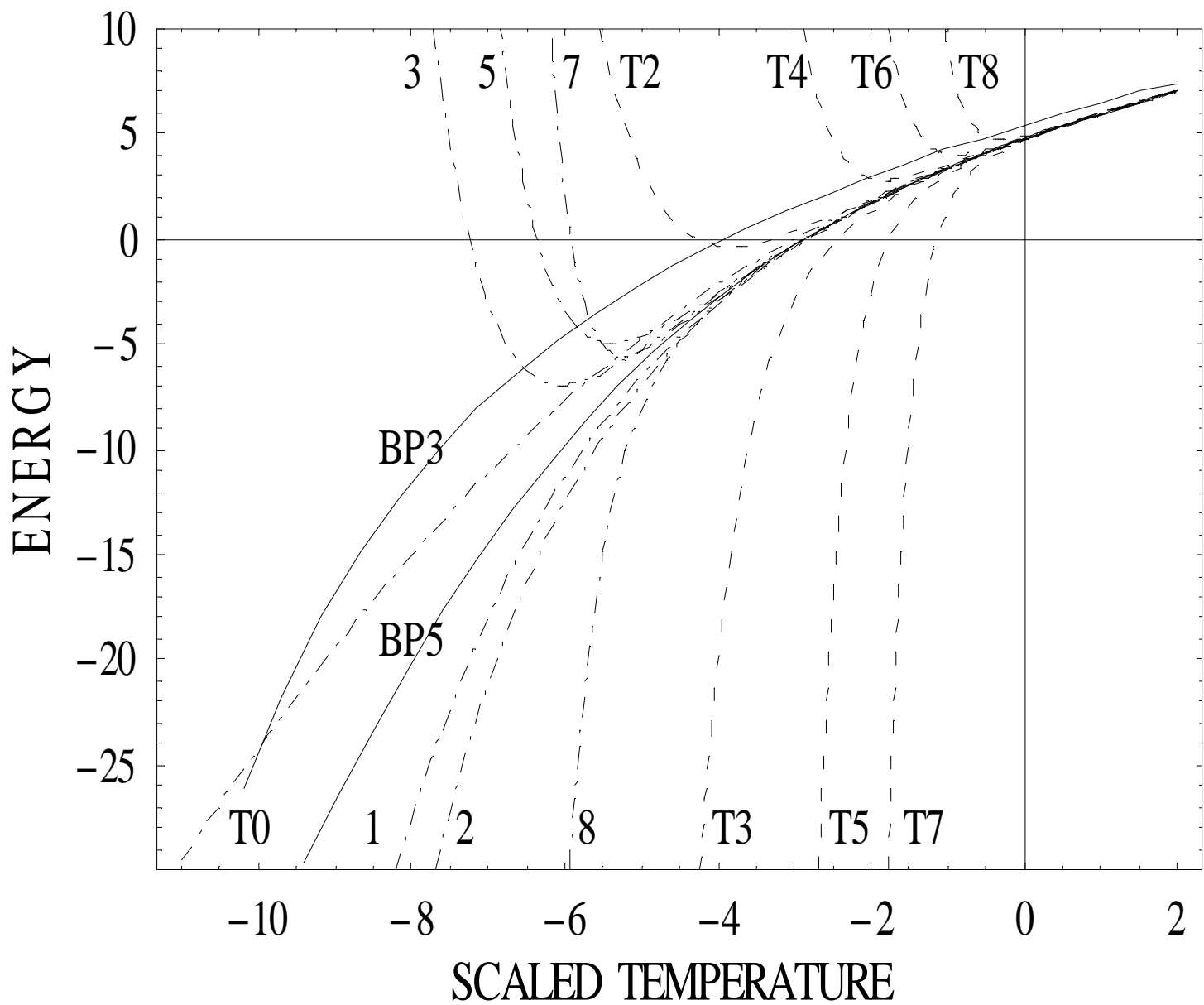
\end{document}